\documentclass[epj,referee]{svjour}
\usepackage{latexsym}
\usepackage{graphics}
\usepackage{graphicx}
\usepackage{amssymb}
\usepackage[dvips]{color}
\begin{document}
\title{Rheology of a sonofluidized granular packing}
\author{Gabriel A. Caballero-Robledo\inst{1,2} \and E. Cl\'ement\inst{2}
}                     
%
%
\institute{Centro de Investigaci\'on en Materiales Avanzados S.C., Nuevo Le\'on, M\'exico. \and Laboratoire de Physique et M\'ecanique des Milieux H\'et\'erog\`enes (UMR 7636 et Universit\'es Paris 6 et Paris 7), 10 rue Vauquelin, 75005 Paris France.}
\date{\today}
%
\abstract{
We report experimental measurements on the rheology of a dry granular material under a weak level of vibration generated by sound injection. First, we measure the drag force exerted on a wire moving in the bulk. We show that when the driving vibration energy is increased, the effective rheology changes drastically: going from a non-linear dynamical friction behavior - weakly increasing with the velocity- up to a linear force-velocity regime. We present a simple heuristic model to account for the vanishing of the stress dynamical threshold at a finite vibration intensity and the onset of a  linear force-velocity behavior. Second, we measure the drag force on spherical intruders when the dragging velocity, the vibration energy, and the diameters are varied. We evidence a so-called ``geometrical hardening" effect for smaller size intruders and a logarithmic hardening effect for the velocity dependence. We show that this last effect is only weakly dependent on the vibration intensity.
} 
\maketitle
\section{Introduction}
\label{intro}
 
A wide class of materials a priori as different as colloidal suspensions, pastes, emulsions, foams or granular materials, show very similar behavior in the vicinity of the jamming transition\cite{LN98}. In all these systems one may observe close to yield many common features such as aging, memory effect or in the fluid phase similar apparent rheology. Recently several propositions have emerged with the aim of describing in a general and unified way this phenomenology (see for example \cite{Nagel02,Bocq09} and refs. inside). For granular matter under shear, the rheology in the dense phase is essentially characterized by a friction coefficient, i.e. a threshold model whose dependence with the shearing rate has been studied extensively \cite{GDR04}. However, in the presence of vibrations much less is known and it is often assumed that the mean kinetic energy should play a role similar to the thermodynamic temperature as for molecular systems. This is essentially the case for granular gases where this analogy can be carried very far. However, this simple vision is currently challenged for dense granular matter. First, for dense granular packing under moderate shearing rates, the velocity fluctuations were shown experimentally and numerically to play a marginal role in the rheology\cite{GDR04}. Second, in the vicinity of the jamming transition as for many other glassy out-of equilibrium phases, novel relations between fluctuation and dissipation were postulated to account for the specific modes of relaxation \cite{MK02}. Many theoretical works suggested to replace the notion of thermal temperature by the more general concept of ``effective temperature(s)" describing the structural evolution between blocked configurations (for a recent review see ref. \cite{Potiguar} and refs. inside). For vibrated granular matter, several attempts were made to identify such relations in numerical simulations \cite{Nicodemi}. But experimentally, for many practical and possibly fundamental reasons, they essentially failed \cite{Zik,Danna} due to the difficultly to probe fluctuations apart from energy input. The issue of approaching jamming can be viewed either from the solid elastic properties (numerically see \cite{OHSLN03} and experimentally see \cite{Tournat,Bonneau}) or from the fluid regime \cite{Sanchez07,Dauchot}; but so far, no consensus has been reached on the fundamental constitutive properties and the rheological behavior of a granular packing under vibration. 

In this report, we access the effective rheology of a weakly vibrated granular packing under a constant confining pressure by measuring the force needed to drag a wire in the bulk at a constant velocity. We observe and characterize the different mobility regimes. To rationalize this situation we propose a simple model that reproduces some of the features observed experimentally. 
In a second series of experiments, we measure the dragging force when the intruder is a bead attached to the wire and we estimate the dragging contribution induced by the presence of the spherical intruder when we vary the diameter and the level of vibration. Note that in the absence of vibration, similar intruder mobility experiments have been carried either in 3D \cite{Zik,Schiffer} or in a 2D granular packing \cite{Behringer}. 
\section{Experimental setup}
\label{sec:setup}
\begin{figure}
\resizebox{\columnwidth}{!}{%
\includegraphics{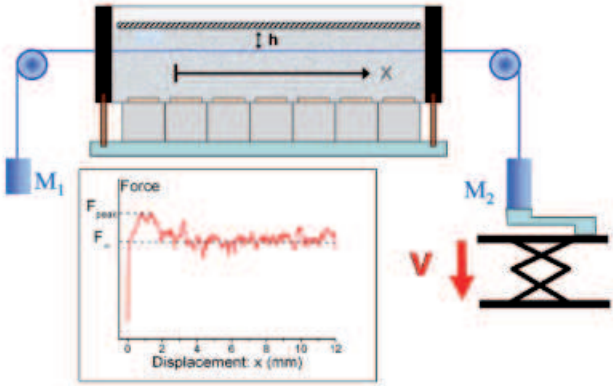}
}
\caption{Sketch of the experimental setup. The driving is at a constant velocity $V$, The masses are such that $M_2>M_1$.  In inset is a typical resistance forces $F(x)$ as a function of the horizontal displacement $x$}
\label{setup}
\end{figure}

The rheomeric device that we use here to explore the effective material rheology is sketched on fig.\ref{setup}. It consists of a glass container closed at its bottom by seven piezoelectric transducers that allow to input in a controlled way, a weak level of vibration in the granular packing. It has been shown by Caballero et al. \cite{Caballero2005} that even a weak level of vibrations may create significant constitutive modifications in the packing: this is the sono-fluidization effect. 

\begin{figure}
\resizebox{\columnwidth}{!}{%
\includegraphics{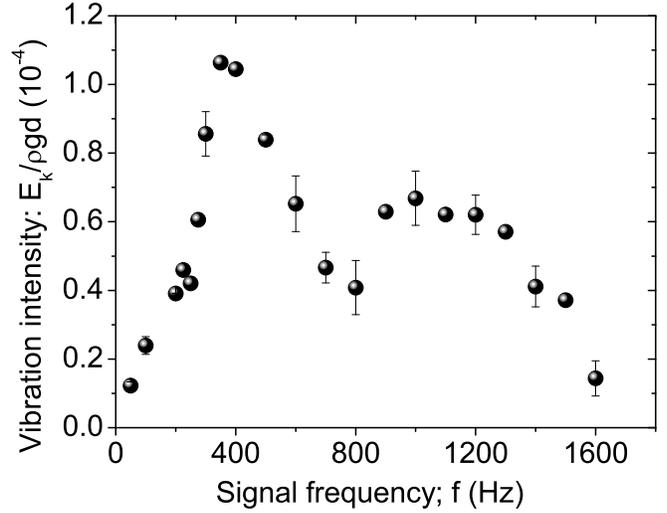}
}
\caption{Agitation energy measured with the accelerometer buried in the bulk as a function of input signal frequency. For these measurements the maximum possible electrical tension was applied to the piezoelectrics.}
\label{fig:Accelerometer}
\end{figure}

The container horizontal section is a rectangle of length $L=19.00(\pm0.05) cm$ and width $W=2.30(\pm0.05) cm$. It is filled with a bidisperse mixture of glass beads of $1.0(\pm 0.1)mm $ and $1.5(\pm 0.2)mm$ mixed in equal masses (i.e. a mean bead diameter of $d=1.2mm$), with the glass density being $\rho=2.6~10^{3}Kg/m^{3}$. This value of bidispersity was chosen to avoid crystallization. In a preliminary version of the experiment, we had noticed that using mono-disperse grains could induce a crystallization which would show up as crystalline patterns visible on the outer surface as well as multiple plateaus in the compaction curve. To us, that was indicating a competing dynamics associated with the growth of crystallites, something we wanted to avoid a priori. Note that we did not perform real systematic measurements on this quite interesting feature. The size and number ratio we use here are just the results on an empirical test
that seemed satisfactory in the context of the present report.   
 The beads are in direct contact with the transducers. Each piezoelectric transducer is excited by a square tension at a frequency $f = 400Hz$ that is set out of phase by $\pi$ compared to that of its neighbors. The external vibration driving is set externally via the electrical tension applied to the piezo transducers. We benchmarked the bulk agitation by monitoring the rms acceleration of a buried accelerometer that indicates a value $\gamma=0.35 g$ ($g$ is the gravity acceleration) for the highest driving voltage. The agitation energy as a function of input signal frequency was measured with the accelerometer buried in the bulk. The results are shown in figure \ref{fig:Accelerometer}. Note that the peak vibration intensity occurs around $400$ Hz, which is the frequency at which the experiments were performed. On the packing top, we placed a thin metallic lid to probe via an induction detector, the surface position. The pressure applied by the lid corresponds to $5$ granular layers: $P_{0}=5 \rho gd=150~Pa$. 

This weak energy input in the granular material does not induce any motion that can be perceived by eye on our current time scale, even if the accelerometer shows that the system is being agitated. 
However, if we follow for a very long time scale the particle motion, we indeed observe substantial reorganization in the granular matrix \cite{Caballero2008}. Note importantly, that we have followed during 10 days, at maximal input energy, a horizontal line of colored beads and we did not evidence any convection motion in the bulk.  

The wire is set in tension by two masses $M_1 < M_2$. The masses values are chosen such that the mass difference will always exceed the dragging force. Therefore, the mass $M_2$ is always in contact with the force probe placed just below. The force probe which is a spring of stiffness $k=10^5 N/m$, is fixed to a vertical translation stage driven downwards at a constant speed, $V$, by a brushless motor. We use two different types of intruders which are dragged horizontally and at a constant depth $h$ below the surface. First, we use a metallic wire of diameter  $\delta=0.1~mm$, spanning the whole horizontal size of the cell, and second, we use the same wire but with a plastic bead of diameter $D >d $ glued to it. The intruders are driven at constant velocities between $V = 10^{-6}~m/s$ and $V= 5~10^{-3} ~m/s$. Hence, the maximal value of the inertial numbers reached for these experiments is $I_{max}\approx \frac{V}{\sqrt{P/\rho}}\approx 10^{-3}$. Hence, the magnitude is much smaller than the values where shear induced hardening would be observed for sheared granular packing \cite{GDR04}.
 
The experiments are conducted using the following protocol. First, the packing and the intruder are vibrated at the maximal vibration energy during 4 hours. We checked in few cases that using 10 hours would not lead to any detectable difference. At the end of this preparation phase a packing fraction of about $\phi = 0.6 $ is reached. Then, the dragging experiment may start at a constant vibration energy characterized by a kinetic energy fluctuation set externally by the voltage on the piezzo transducers. A typical force signal is presented in inset of figure (\ref{setup}). Note that at the beginning of the pulling we observe an overshoot of the resisting force, but after few millimeters of displacement a steady-state is reached.

\section{Drag force on the wire}
\label{sec:drag}
\begin{figure}
\resizebox{\columnwidth}{!}{%
\includegraphics{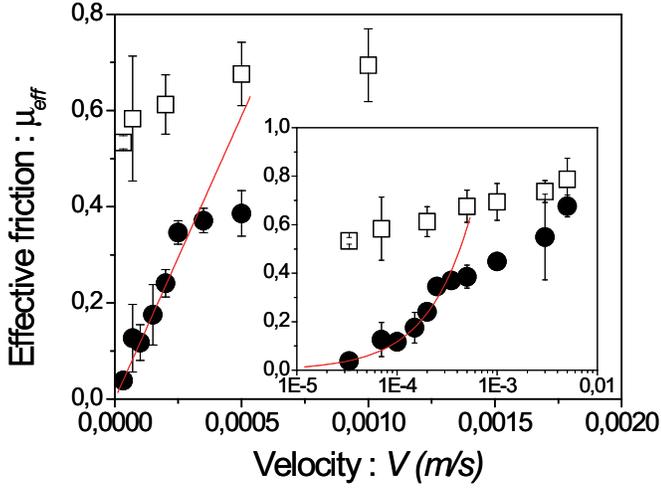}
}
\caption{Effective friction on the wire intruder $\mu_{eff}$ as a function of the pulling velocity $V$;  no vibration ($\square$), maximal vibration ($\bullet$). In inset is the same graph but is a lin-log scale. The red line is the curve : $y=x/8.5~10^{-4}$.
}
\label{dragthread}
\end{figure}

In a first series of experiments the effective rheology of the packing is probed by monitoring the drag force on a metallic wire. The wire is placed at a height $h=1.3 cm$ below the surface. The average force necessary to drag the wire $<F_{t}>$, is rescaled by the wire outer surface ($S_{t}=6~10^{-5}m^2$) in order to obtain an average friction stress : $\sigma_{t}=<F_{t}>/S_{t}$. Then, the average friction stress is rescaled by the confining pressure $P(h)=P_{0}+\phi \rho_{s} g h$ to form an effective friction coefficient: $\mu_{eff}=\sigma_{t}/P(h)$. This effective friction is displayed on fig.(\ref{dragthread}) as a function of the dragging velocity. We use two representations, a linear scale to display the lower velocity part of the measurements and in inset, a linear-log scale representing the whole range of measurements. The error bar corresponds to a r.m.s. deviation over 5 independent experiments. The data show several salient and robust features. First, at a constant driving velocity and for a small level of vibration, the pulling resistance decreases strongly with the vibration energy. Second, with or without vibration, the stress-strain relations are of the hardening type: we indeed observe an increase of the resistance to pulling with the dragging velocity. The third striking feature is that for the highest vibration energy and the lower driving velocities, we reach a linear regime of the type: $\mu_{eff}\approx V/V^{*}$ with $V^{*} \simeq 8.5 10^{-4}m/s$. Finally, for weak vibration energy, the increase of friction with the driving velocity is weak, typically of the type $\mu = \mu_{0} + \beta \ln (V/V_{0})$, where $V_{0}$ is a reference velocity. This behavior is also observed in the absence of vibration on about two decades in velocity. However, in this last case, we could not identify clearly the existence of a dynamical friction threshold as a well defined plateau.

The inertial number being quite weak as noticed before, the hardening effect observed here cannot be attributed to a shear induced hardening (at least in the usual sense as described in \cite{GDR04}). It is possible that the weak increase of friction with velocity comes from an intrinsic solid on solid friction property at the level of the granular contacts. Note that such a behavior was evidenced for granular packing undergoing very slow shear against a wall \cite {Ovarlez03}.

\section{A drag force model}
\label{sec:model}

\begin{figure*}
\begin{center}
\includegraphics[scale=0.6]{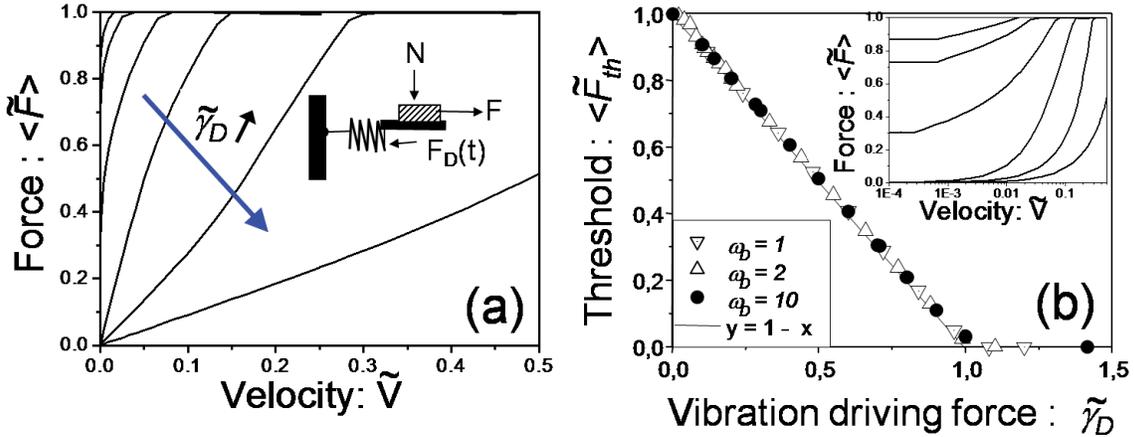}
\end{center}
\caption{Rheology of the drag resistance model, coupling solid friction and elasticity under harmonic driving (see sketch on inset of \ref{model}a). On figure \ref{model}a is the rescaled average drag force $<\widetilde{F}>$ as a function of the rescaled pulling velocity $\widetilde{V}$ for an harmonic driving at $\widetilde{\omega }_{d}=10$ and for different vibration forcing amplitudes $\widetilde{\gamma_{d}} =0.28$(leftmost), $0.4$, $0.7$, $1.4$, $2.8$, and $7$(rightmost). On figure \ref{model}(b) is the low velocity drag force threshold $<\widetilde{F}_{th}>$ versus the vibration forcing amplitudes $\widetilde{\gamma_{d}}$ for driving frequencies $\widetilde{\omega }_{d}=1, 2, 10$. The data collapse on the curve $<\widetilde{F}_{th}> = 1 - \widetilde{\gamma_{d}}$ (straight line). In inset, is displayed  figure \ref{model}(a) in a lin-log scale to evidence the low velocity dynamical force threshold}
\label{model}
\end{figure*}

Here we present a simple heuristic model suited to rationalize the passage from a threshold rheology of the Coulomb type, to a linear force-velocity relation. The model shows explicitly that in the framework of solid friction coupled with elasticity, one can perfectly understand this change of rheology and the disappearance of a Coulomb threshold due to external vibration. Note that many sophisticated variants of such an heuristic spring/friction model exist in the literature, essentially to account for the complex stick-slip dynamics of earth-quakes \cite{Knopoff}, or to explain granular rheology fluctuations \cite{Behringer}. Under external forcing, similar models were proposed in the engineering context to render the fatigue of structures \cite{Karnopp66,Bouc98}. In the presence of shearing vibration, the solid/solid friction of a block on an inclined plane was investigated via a spring/friction model with memory \cite{Bureau} but the emphasis was rather on the stick-slip transition and the influence of mechanical vibrations on thermal aging. Note that it could be interesting in a future work, to bridge all these approaches and extend the present model to account for granular reorganization as well as contact thermal aging. A similar model was presented recently to account for the effect of a horizontally vibrating base on a spreading granular layer \cite{Sanchez07}.

A scheme of the model is sketched as an inset in figure \ref{model}(a). It consists of a linear spring of mass $M$ in frictional contact with a block. On the block a constant normal force $N$ is applied. The interaction between the spring and the block is through solid-solid friction characterized by a Coulomb friction coefficient $\mu$, while the spring elasticity is characterized by a resonant angular frequency $\omega_{0}$. A force $F(t)$ acts on the block such as to drive it at a constant velocity $V$. In analogy with the experiment, the block represents the moving intruder while the spring accounts for the elastic response of the granular medium. The normal force $N$ results from the confinement pressure exerted by the medium on the intruder, and the friction coefficient accounts for the friction interactions. 

To render the influence of the piezo transducers exciting the internal elastic modes of the packing, we use a harmonic external driving force $F_{d}(t)=\mu N~\widetilde{\gamma_{d}}\sin \left( \omega t \right)$ (see fig. \ref{model}), acting on the spring; the rescaled driving frequency is $\widetilde{\omega }_{d}=\frac{\omega}{\omega_{0}}$, the non dimensional time is $\widetilde t = \omega_{0}t$ and  $\widetilde{\gamma_{d}}$ is the vibration force amplitude rescaled by the friction force $F_0 = \mu N$ \footnote{In the rest of the document, the tilde signs refer to rescaled non dimensionalized quantities.}. 
	
When the block and the spring stick together, we have $\dot{\widetilde{x}} = \widetilde{V}$, where ${\widetilde{x}} = \frac {xM \omega_{0}^{2}}{\mu N}$ is the non dimensional position of the spring and $\widetilde{V} = \frac{VM \omega_{0}}{\mu N}$ is the non dimensional velocity of the block. In this case, the equation of motion is :
\begin{equation} 
\widetilde{F} = \widetilde{x} + \widetilde{F_{d}}, 
\label{eq:stick}
\end{equation}
as long as $\left| \widetilde{F} \right| <1$, with the forces $F$ and  $F_d$ rescaled by $F_0 = \mu N$. 

When $\left| \widetilde{F} \right| = 1$ there is a relative sliding velocity between the spring and the block.  In this case the spring equation of motion is:
\begin{equation} 
\ddot{\widetilde{x}}=-\widetilde{x}+\widetilde{F}-\widetilde{F_{d}},
\label{motion}
\end{equation}
with $\widetilde{F_{d}}$ the external driving force, and 
\begin{equation} 
\widetilde{F}=sgn(\widetilde{V}-\dot{\widetilde{x}}).
\label{force}
\end{equation}
Note that for simplicity, we do not include here any internal dissipation mechanism that would damp the elastic resonance. 
 
For large vibrations, $\left| \widetilde{F}\right|<<\left| \widetilde{F_{d}}\right| $, the block and the spring always slide on each other (no sticking). In this case, equation (\ref{motion}) reduces to the equation of a harmonic oscillator driven by the external force $\widetilde{F_{d}}$. It can then be easily solved. The velocity is then : 
\begin{equation} 
\dot{\widetilde{x}}=-\frac{\widetilde{\omega }_{d}}{\widetilde{\omega }_{d}^{2}-1} ~ \widetilde{\gamma _{d}}  \cos \widetilde{\omega }_{d} \widetilde{t},
\label{velocity}
\end{equation}
and the the r.m.s. spring velocity is :
\begin{equation}
\widetilde{\overline{V_s}} \equiv \left\langle  \dot{ \widetilde{x} }^2 \right\rangle ^{1/2} = \frac{\sqrt{2}}{2}\frac{\widetilde{\omega }_{d}}{\left| \widetilde{\omega }_{d}^{2}-1\right| } \widetilde{\gamma _{d}}.
\label{eq:rms}
\end{equation}
Considering that the block is moving at constant velocity $\widetilde{V}$, one can derive from equations (\ref{force}) and (\ref{velocity}) the low velocity limiting value, $\widetilde{V} \rightarrow 0$, of the average friction force, characterized by an effective friction coefficient:
\begin{equation} 
 \mu_{eff} \equiv \frac{\left\langle F \right\rangle}{N} = \mu \left\langle \widetilde{F} \right\rangle =
 \frac{2\widetilde{V}}{\pi}\left( \frac{\widetilde{\omega }_{d}}{ \widetilde{\omega }_{d}^{2}-1} \widetilde{\gamma _{d}}\right) ^{-1},
\label{drag}
\end{equation}
which in terms of $\widetilde{\overline{V_s}}$, eq. \ref{eq:rms}, can be written as
\begin{equation} 
 \mu_{eff} =  \mu \frac{\sqrt{2}}{\pi} \frac{V}{\overline{V_s}}.
\label{drag2}
\end{equation}
 Equation (\ref{drag2}) states that under strong external vibration and a very low velocity driving of the block, the effective friction force that resists to motion increases linearly with the driving velocity. In this limit, the only important parameter that characterizes vibration is the r.m.s. vibration velocity. That is quite consistent with what is observed experimentally under such conditions, as it is shown on figure \ref{dragthread}. By numerically solving eq. \ref{motion}, we display on fig. \ref{model}(a) the rescaled average drag force $ \left\langle \widetilde{F} \right\rangle = \frac{ \left\langle F\right\rangle }{\mu N}$ as a function of the rescaled velocity $\widetilde{V}$ for different vibration amplitudes (labeled here in terms of the rescaled vibration force amplitude $\widetilde{ \gamma_{d} }$), at a constant driving frequency $\widetilde{\omega }_{d}=10$. We see that when the amplitude is increased, one obtains for $\gamma_{d} \approx 1 $, a transition to a linear force velocity regime characterized by the limiting force-velocity relation of equation \ref{drag}. The validity of limiting analytical solution (equation(\ref{drag2})) was verified explicitly by comparison with the numerical solution. Another interesting feature is that this model displays a transition between a threshold rheology at low vibrational driving, to a linear force/velocity rheology at higher vibration amplitudes.
In the specific context of this model, the effective friction threshold to trigger relative sliding can be written : 
\begin{equation}
\mu_{thresh}= \mu (1 - \widetilde{ \gamma_{d} } ). 
\label{eq:muthresh}
\end{equation}
Thus the threshold vanishes for $\widetilde{ \gamma_{d} } \geq 1$ (see fig.\ref{model}(b)). Relation (\ref{eq:muthresh}) can be simply understood by inspection of equation (\ref{eq:stick}), where we see that the condition $\left| \widetilde{F} \right| <1$ is valid as long as $\widetilde{x}$ is smaller than $\widetilde{x_M} = 1 - \widetilde{ \gamma_{d} } $, which immediately yields the value for the effective friction threshold (eq. \ref{eq:muthresh}). This is in essence what seems to happen experimentally. It would be interesting in further experimental work, to test more thoroughly the generic existence of this type of transition between a threshold fluid behavior and a linear fluid behavior for vibrated packing. 
At this point, we do not seek to match exactly the experimental results, first because experimentally we do not have a clear vision of which elastic modes are the most strongly coupled to the wire friction. Also, because this model is not suited to explain the slow (logarithmic) variations of the dragging force. This last behavior may be due to intrinsic contact aging properties or unpinning dynamics in a disordered environment. We just notice that the rms velocity $\overline{V_s} \simeq 1.2~10^{-3} m/s$ measured for the largest available vibration is of the order of the velocity $V^{*}=8.5~10^{-4} m/s$ found for the effective friction relation in the linear regime (fig. \ref{dragthread}). This is quite consistent in magnitude with the proposed theoretical picture.
Finally, we just comment that obtaining the elastic stiffness of the effective spring that is actually coupled to the wire, is a complicated issue. However, we can provide an order of magnitude calculation. Recent experimental measurements on surface waves in a granular packing under gravity have shown that under such small confinement pressure, the propagation velocity for wavelengths of the order of one centimeter can be as low as $c=40~m/s$ \cite{Tournat}, \cite{Bonneau}, where $c \approx \sqrt{G/\rho}$, $G$ being an effective shear stiffness. If we estimate the effective spring stiffness to be $k \approx G L W/h$, and the mass set into motion as $M = \rho h W L$, we  get a resonance angular frequency $\omega_0 = \sqrt{k / M} = c / h$. The order of magnitude calculation of the resonance frequency $f_0 \approx 500~Hz$, is about what we observe experimentally (see fig. \ref{fig:Accelerometer}).

\section{Drag force on an intruding bead}
\label{sec:intruder}

\begin{figure*}[t!]
\includegraphics[width=170mm,keepaspectratio=true]{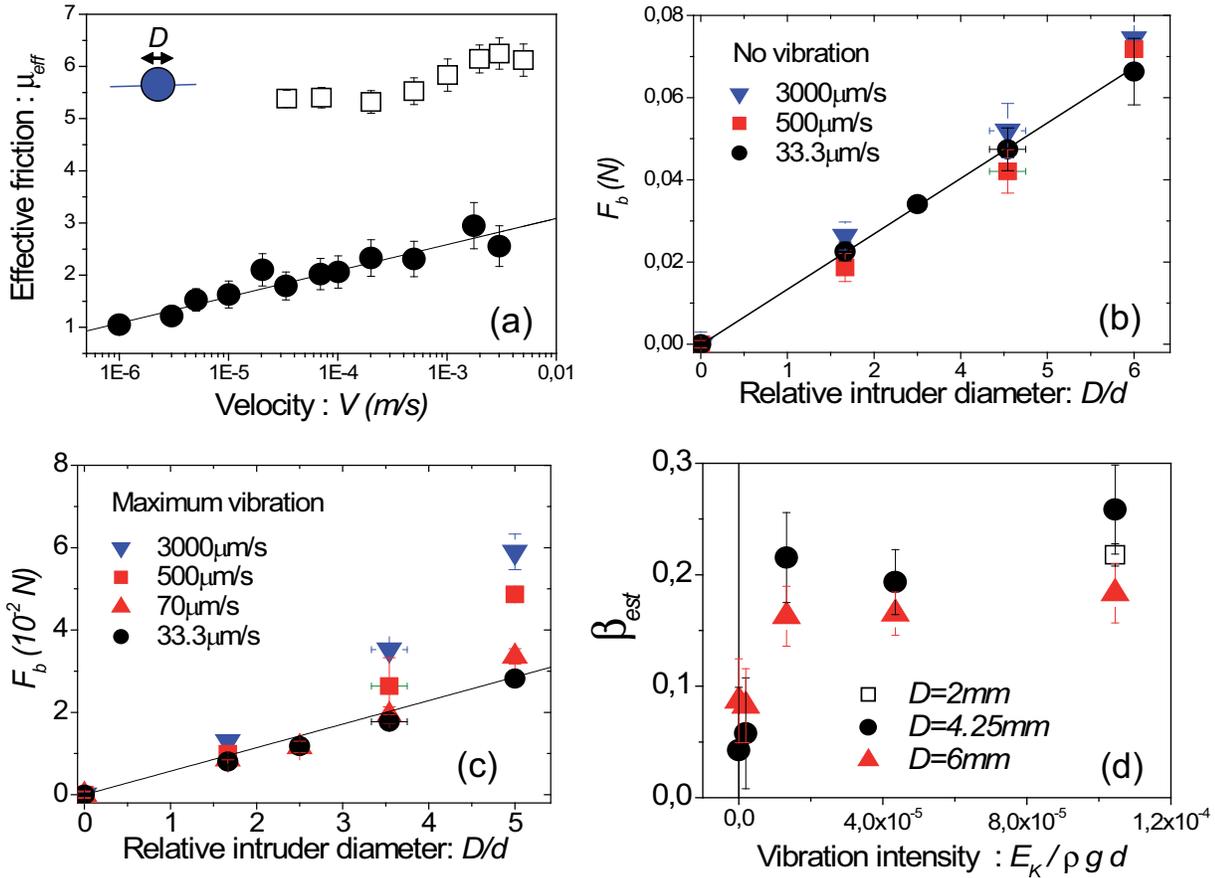}
\caption{Rheological properties of an intruding bead of diameter $D$. (a) Effective friction for an intruder of diameter diameter $D =2mm$ as a function of the driving velocity, for no vibration ($\square$) and at the largest accessible vibration energy ($\bullet$). (b) Dependence of the bead drag force contribution $F_{b}$ as a function of size ratio $D/d$ in the absence of vibration and (c) for the maximal vibration. (d) Estimation of the logarithmic hardening slope $\beta_{est}$ as a function of the vibration kinetic energy for different diameters.}
\label{dragbead}
\end{figure*}

In this second series of experiments, we measure the drag force exerted on an spherical intruder of diameter larger than the grains. An essential difference with the previous experiment is that now, the intruder motion induces a back flow and thus, granular reorganizations in the bulk. Pulling a spherical intruder at a constant velocity in the bulk of a granular packing presents many practical difficulties. Here, we try to override this problem by considering the resisting force on an intruder with a bead attached to the metallic wire. The wire (with the bead attached) is put originally at an horizontal position $h=1.3~cm$ below the surface and the resisting force on the system is measured. To extract the contribution of the bead $F_{b}$, we also measure the drag force on the wire only in an independent experiment at the same velocity and remove the corresponding mean resisting force from the drag force on the wire+bead intruder. Then, to obtain an effective friction coefficient that characterizes the bead-only contribution, we divide $F_{b}$ by the bead surface and by the confining pressure $P$ : $\mu_{eff}= F_{b}/\pi D^2/(\phi \rho g h +P_{0})$. This simple subtraction procedure may be questionable as, in principle, both contributions cannot be as easily decoupled. The presence of the bead may modify the overall friction stress on the wire with respect to the situation where the wire only is present. We nevertheless expect that this effect will be small as the spatial extend of the perturbation due to the presence of the bead remains small compared to the wire length. The corresponding effective frictions are displayed on fig. \ref{dragbead}(a) for a intruding bead of diameter $d =2~mm$. The error bars correspond to a typical r.m.s. dispersion over 3 independent experiments.

We noticed that before reaching a steady state, typically by pulling the intruder over a distance corresponding to a mean grain size $d$ (this, almost independently of the intruder size), we go through a phase where an overshoot of the dragging force is systematically evidenced. Also, similarly to the wire only case, the rheology is of the hardening type  and the higher is the vibration, the lower is the resistance to pulling. 

In the absence of vibration we apparently reach, at low driving velocities, a threshold value smaller than its limiting value at higher velocities. Note that in previous dragging experiments performed in 2D, Geng et al. \cite{Behringer} observe a very similar phenomenology. We estimate in this last case, the typical dragging velocities as being around $10^{-4} m/s$. The confining pressure being of the order of the drag force in their experiment, they had an effective inertial number around $10^{-4}$, which yields working conditions and a phenomenology consistent with the present report. But either in their case or in ours, the reason for hardening effect observed at such low velocities is unclear. We cannot dismiss, as in the case of the wire only, the existence of an intrinsic contribution coming from aging of granular contacts. However, note that in previous numerical works \cite{daCruz05,Aronson08}, it has been shown that material softness could also play a role as it leads to a hardening regime decoupled from the shear hardening regime for hard spheres. For the 3D dragging experiment of a rod by Albert et al. \cite{Schiffer}, no hardening effect was evidenced but the dragging velocities were essentially higher than our operating conditions.
 
Interestingly and contrarily to the wire-only case, when the vibration is activated up to its maximal available value, we never could evidence a linear force-velocity regime, and this for any of the intruders we tested. We found essentially a slow logarithmic increase of the friction as displayed on fig.\ref{dragbead}a with $D=2mm$ on 4 decades in velocities. 

Now we vary the intruder size $D$. On fig. \ref{dragbead}b and c we display, for different dragging velocities, the resisting forces without vibration (see fig.\ref{dragbead}b) and for the maximal vibration (see fig. \ref{dragbead}c) at different dragging velocities. The choice to represent directly the force instead of the effective friction is motivated by the fact that the dragging force varies linearly with the intruding grains diameter $D$. 
Note that at the highest vibration level and for the highest drag velocities, we observe significant deviations from linearity. This linear $F_{b}\propto D$ relation is even more surprising if we choose to represent it as an effective friction coefficient. Then, the drag force would be divided by the bead intruder surface ($\propto D^2$) and we obtain a striking ``geometrical hardening" effect for the smaller intruder sizes ($\mu_{eff}\propto D^{-1}$). Interestingly, looking back at the data of Albert et al. \cite{Schiffer}, we indeed observe for the drag force on the cylinder, a crossover between a linear behavior (here corresponding to a constant effective friction) at large intruder sizes, and a plateau corresponding to the ``hardening" effect for smaller grain sizes. However, this effect is not present in the data of Geng et al. \cite{Behringer}. 

Note that in the absence of vibration, the effective stress values are:  $\mu_{eff}(d=2~mm)=6$ and $\mu_{eff}(d=6~mm)=2$. It is then straightforward to calculate theoretically the friction force exerted on the surface of a sphere dragged in a medium for which the confining pressure is hydrostatic. If the local dynamical friction is $\mu$, the effective friction would then be: $\mu_{eff}=\frac{\pi}{2} \mu$. Therefore, the actual values for the effective friction coefficients found experimentally are quite large when compared to the usual values found for cohesionless granular packing. This means that the effective friction should take into account a flow processes in the bulk which involves many grains flowing around the intruder. It is clear that in this case, the lateral confinement could also play a role in the determination of the effective rheology, but still this does not explain why the effective friction would be enhanced for smaller grains sizes. 

Finally, we estimate the dependence of the logarithmic hardening effect with the vibration energy. We seek to estimate the logarithmic slope of the effective rheology : $\beta =\partial \mu _{eff}/\partial \ln V$. In the context of solid on solid friction, this is a parameter quite sensitive to temperature which value increases significantly in the vicinity of the glassy transition \cite{Baumberger99}. Note that the whole protocol to determine a full rheological curve as in fig. \ref{dragthread}, is quite cumbersome and lengthy. It took typically three months worth work. This explains why these results may look fragmented as we have not performed yet extensive and systematic measurements for a large range of driving velocities and vibration intensities. We rather choose to estimate the value of $\beta$ and the impact of vibration on its value. We actually performed two series of experiments at well separated driving velocities $V = 3.3~10^{-5}m/s$ and $V= 5~10^{-4}m/s$. From those two points which average is determined over 5 independent realizations, we estimate the logarithmic slope $\beta_{est}$. The results are displayed on fig.\ref{dragbead}d. We see that, indeed, the increase of the vibration energy by a factor 5 does not influence very much the velocity hardening effect characterized by $\beta_{est} $. It is possible however from the data, that in the absence of vibration the hardening effect would be smaller (typically a factor 2).
\section{Summary and conclusion}
\label{sec:summary}

In this report we presented experimental results demonstrating that even for a weak level of vibration, significant changes can be evidenced for the dynamical yield stress and the rheology of a granular packing. Two types of solid intruders pulled at constant velocity are used and we show that injecting sound waves in the bulk leads to important drag reductions.  The resisting force generally increases with the pulling velocity, thus the rheology is of the hardening type. Interestingly, when a wire alone is driven very slowly, we observe a transition from a yield stress rheology characterized by an effective friction increasing slowly with velocity, to a linear force-velocity regime where the threshold vanishes for a finite energy input. The values obtained for the effective friction coefficient are consistent with a simple heuristic model coupling internal elasticity with solid friction.  The logarithmic hardening regime obtained in the absence of vibration could be due either to vibrations generated internally by the bead friction mechanism or, alternatively, by the intrinsic granular contact properties as evidenced by Ovarlez et al.\cite {Ovarlez03}. Interestingly, when a wire alone is driven very slowly, we observe a transition from a yield stress rheology characterized by an effective friction increasing slowly with velocity, to a linear force-velocity regime where the threshold vanishes for a finite energy input. The values obtained for the effective friction coefficient are consistent with a simple heuristic model coupling internal elasticity with solid friction.  The logarithmic hardening regime obtained in the absence of vibration could be due either to vibrations generated internally by the bead friction mechanism or, alternatively, by intrinsic granular contact properties as evidenced by Ovarlez et al.\cite {Ovarlez03}. We also measured the contributions to the drag force of spherical intruders of various sizes which motion induces a back-flow in the granular packing. In this case, logarithmic force-velocity relations were observed. At a constant drag velocity, we obtain a linear increase of the resisting force with the intruder size corresponding to a ``geometrical hardening" effect that enhances the effective friction for smaller intruders. This is observed both in the absence of external vibration and at the maximal level of vibration. Finally, we estimate the logarithmic velocity hardening for different levels of vibration and found that it is only marginally influenced by the actual level of kinetic energy. These two last effects remain a central issue to the understanding of granular packing rheology with and without vibration.

\begin{acknowledgement}
We thank Jos\'e Lanuza for help on the experimental device and Pr. Bruno Andreotti for discussion and critical reading of the manuscript.
\end{acknowledgement}


\begin{thebibliography}{}
%
%


\bibitem{LN98} A.Liu, S.R.Nagel, Nature {\bf 396}, (1998) 21.
\bibitem{GDR04}GDR Midi(a collective work), Euro.Phys.J E \textbf{14}, (2004) 341.
\bibitem{Nagel02} C.O'Hern, S.langer, A.Liu,S.R.Nagel Phys. Rev. Lett. \textbf{88}, 036001 (2002).
\bibitem{Bocq09} L.Bocquet, Phys. Rev. Lett. \textbf{103}, 036001 (2009).
\bibitem{MK02} H.A. Makse and J. Kurchan, Nature {\bf 415}, (2002) 614.
\bibitem{Potiguar} F. Q. Potiguar and H. A. Makse, Eur. Phys. J. E \textbf{19}, (2006) 171.
\bibitem{Nicodemi}  M.Ciamarra, A.Coniglio, and M.Nicodemi, Phys. Rev. Lett. \textbf{97}, (2006) 158001.
\bibitem{Zik} O.Zik, J.Stavans J and Y.Rabin, Europhys. Lett. {\bf 17}, (1992) 315.
\bibitem{Danna} G.d'Anna et al. Nature \textbf{424}, (2003) 909.
\bibitem{Sanchez07} I. Sanchez et al., Phys.Rev.E \textbf{76}, (2007) 060301(R).
\bibitem{Dauchot}F. Lechenault, O. Dauchot, G. Biroli, J.-P. Bouchaud, Europhys.Lett.\textbf{ 83}, 46003 (2008).
\bibitem{OHSLN03} C. S. O'Hern, L. E. Silbert, A. J. Liu and S. R. Nagel, Phys. Rev. E \textbf{68}, (2003) 011306.
\bibitem{Tournat}V. E. Gusev, V. Aleshin, and V. Tournat, Phys.Rev.Lett. \textbf{100}, (2008) 158003.
\bibitem{Bonneau} L. Bonneau, B. Andreotti and E. Clement, Phys.Rev.Lett. \textbf{101}, (2008) 118001.
\bibitem{Caballero2005} G.Caballero, J.Lanuza and E.Clement, \textit{Powders and Grains 2005}, eds Garcia-Rojo, H.A. Herrmann and S.McNamara (Taylor and Francis, 2005), p339.
\bibitem{Schiffer} R. Albert et al., Phys. Rev. Lett. {\bf 82}, 205.(1999); I. Albert et al. Phys.Rev.Let. {\bf 84}, (2000) 5122.
\bibitem{Behringer}J. Geng and R.P. Behringer, Phys. Rev. Lett. \textbf{93}, 238002 (2004); Phys. Rev. E \textbf{71}, (2005) 011302 .
\bibitem{Caballero2008}G.Caballero, C.Goldenberg and E.Clement, in preparation (2009)
\bibitem{Knopoff}Burridge, R. and Knopoff, L. Bull. Seis. Soc. Amer. \textbf{57}, (1967) 341.
\bibitem{Bureau}L. Bureau, T. Baumberger, and C. Caroli, Phys. Rev. E \textbf{64}, (2001) 031502. 
\bibitem{Karnopp66} D.Karnopp, T.D. Scharton , J.of Acoust.Soc.Am. \textbf{39}, 1554 (1966).
\bibitem{Bouc98} R.Bouc, D.Boussa, C.R. Acad. Sci. Paris \textbf{326},Srie II, 475 (1998).
\bibitem{daCruz05}F. da Cruz et al., Phys. Rev. E \textbf{72}, (2005) 021309.
\bibitem{Aronson08}I.S. Aranson et al., Phys. Rev. E \textbf{78}, (2008) 031303.
\bibitem{Baumberger99} T.Baumberger, P.Berthoud and C.Caroli, Phys. Rev. B \textbf{60}, (1999) 3928.
\bibitem{Ovarlez03} G.Ovarlez, E.Clement, Phys. Rev. E \textbf{68}, (2003) 031302.
\end{thebibliography}
%

\end{document}